\documentclass[preprint, 5p, times,twocolumn ]{elsarticle}

\usepackage{amsbsy,amssymb,amsmath,bm}
\usepackage{graphicx,color,epsfig,rotate}

\journal{Journal of Crystal Growth}

\begin{document}

\begin{frontmatter}

\title{Single Crystal Growth of Ga$_2$(Se$_x$Te$_1$$_-$$_x$)$_3$ Semiconductors and Defect Studies via Positron Annihilation Spectroscopy}

\author{N.M. Abdul-Jabbar$^1$$^,$$^2$, E.D. Bourret-Courchesne$^2$, B.D. Wirth$^1$$^,$$^3$}

\address{$^1$Department of Nuclear Engineering, University of California, Berkeley, California 94720, USA}
\address{$^2$Materials Sciences Division, Lawrence Berkeley National Laboratory, Berkeley, California 94720, USA}
\address{$^3$Department of Nuclear Engineering, University of Tennessee, Knoxville, Tennessee 37996, USA}

\begin{abstract}
Small single crystals of Ga$_2$(Se$_x$Te$_1$$_-$$_x$)$_3$ semiconductors, for \emph{x} = 0.1, 0.2, 0.3, were obtained via modified Bridgman growth techniques. High-resolution powder x-ray diffractometry confirms a zincblende cubic structure, with additional satellite peaks observed near the (111) Bragg line. This suggests the presence of ordered vacancy planes along the [111] direction that have been previously observed in Ga$_2$Te$_3$. Defect studies via positron annihilation spectroscopy show an average positron lifetime of $\approx$400 ps in bulk as-grown specimens. Such a large lifetime suggests that the positron annihilation sites in these materials are dominated by defects. Moreover, analyzing the electron momenta via coincidence Doppler broadening measurements suggests a strong presence of large open-volume defects, likely to be vacancy clusters or voids.       

\end{abstract}

\begin{keyword}
Ga$_2$(Se$_x$Te$_1$$_-$$_x$)$_3$, Structural vacancies, Positron

\end{keyword}

\end{frontmatter}

\section{Introduction}

Ga$_2$(Se$_x$Te$_{1-x}$)$_3$ semiconductors belong to a class of III-VI materials that exhibit a cubic zincblende crystal structure (F\={4}3\emph{m} space group) dominated by stoichiometric or ``structural'' vacancies. These structural defects arise due to the inherent valence mismatch between the anion and cation forcing 1/3 of the cation sites to be vacant. Previous investigations have posited that the presence of such stoichiometric defects in these classes of materials can greatly influence their material properties. Studies on the electrical properties of these class of materials show a rapid increase in charge carrier mobility as a function of temperature. For instance, a maximum Hall mobility of 210 cm$^2$/Vs at $77\,^{\circ}\mathrm{C}$ for p-type In$_2$Te$_3$ has been measured \cite{Sen_1984}. Similar behavior has been seen in p-type Ga$_2$Se$_3$ \cite{Gamal_1995, Belal_1995}.

Diffraction and electron microscopy studies on Ga$_2$Te$_3$ have revealed an ordered-vacancy orthorhombic superlattice in addition to the nominal disordered zinc blend structure \cite{Newman_1963, Chou_1993, Hutchison_1994}. X-ray diffraction studies on Ga$_2$Se$_3$ have shown the presence of a monoclinic phase with zigzag ordered vacancies \cite{Lubbers_1982}. Theoretical work on the lattice dynamics of ordered defect zincblende structures suggests that the presence of ordered vacancies does not strongly affect crystal lattice vibrational modes \cite{Finkman_1973, Finkman_1975}. Nevertheless, the precise role that ordered structural vacancies have on material electrical properties remains uncertain.

From a purely structural perspective, III-VI defect zincblende semiconductors display anomalously high radiation stability. Single crystal and polycrystalline specimens of In$_2$Te$_3$, Ga$_2$Te$_3$, and Ga$_2$Se$_3$ were exposed to a flux of 1.2 MeV $\gamma$-quanta up to a dose of 3 $\times$ 10$^{22}$ m$^{-2}$, a flux of fast electrons (with energies up 100 MeV) up to a dose of 3 $\times$ 10$^{23}$ m$^{-2}$, and a flux of mixed $\gamma$-neutron  radiation from stationary and pulsed nuclear reactors up to a dose of 10$^{23}$ m$^{-2}$ \cite{Koshkin_1994}. Parameters such as charge carrier concentration, charge carrier mobility, and microhardness measured before and after irradiation show little or no change \cite{Koshkin_1994}. Such observations can be explained by the presence of structural vacancies which minimize Frenkel pair production from incident radiation (as opposed to elemental and III-V cubic semiconductors with no fractional cation occupancies) \cite{Koshkin_1994, Gurevich_1995}. As a result, these class of materials may be applicable for nuclear particle detection or for semiconductor devices operating in high radiation environments.  

The aim of this work is to evaluate the potential of defect zincblende III-VI semiconductors for room-temperature $\gamma$-ray spectroscopy applications and investigate the effects of their intrinsic microstructural material properties on device performance (where the main requirement for semiconductor radiation detectors is large $\mu$$\tau$ values). Ga$_2$Te$_3$ is taken as a starting point, since it lacks the high to low temperature phase transitions observed in In$_2$Te$_3$ and  Ga$_2$Se$_3$ which can make large single crystal growth difficult \cite{ASM}. A band gap of 1.1 eV and an electron mobility of 28 cm$^2$/Vs at $37\,^{\circ}\mathrm{C}$ for Ga$_2$Te$_3$ have been reported \cite{Sen_1984}. Selenium is then alloyed to the binary compound to produce the ternary Ga$_2$(Se$_x$Te$_{1-x}$)$_3$ system in an effort to engineer the band gap for room-temperature radiation detector applications and to study the effect of selenium addition on the material properties. This paper presents initial structure and defect investigations for this class of material.     

\section{Synthesis and Crystal Growth}

Ga$_2$(Se$_x$Te$_{1-x}$)$_3$ compounds were synthesized using stoichiometric amounts of 8N gallium, 6N selenium, and 6N tellurium sealed in quartz crucibles (1.3 cm in diameter) under vacuum (10$^{-5}$ to 10$^{-6}$ Torr). The quartz crucible was placed on an incline (to allow for good mixing) in an alumina boat, and the boat was placed in a 20 cm horizontal tube furnace. Figure 1\emph{a} shows a schematic of the setup in the furnace. A synthesis temperature of $850\,^{\circ}\mathrm{C}$ was used, where the measured profile showed a gradual rise in temperature going from the left to the right of the furnace. Specifically, temperatures of $830\,^{\circ}\mathrm{C}$ at the left end, $855\,^{\circ}\mathrm{C}$ in the middle, and $860\,^{\circ}\mathrm{C}$ at the right end were measured. The stoichiometric amounts of gallium, selenium, and tellurium were allowed to react for 36 h, then the system was cooled directionally at a rate of $0.3\,^{\circ}\mathrm{C}$/min in an attempt to grow small single crystals. Resulting ingots were composed of small single crystals mixed in a polycrystalline mass of the same composition. The single crystal were harvested (Figure 1\emph{b}) and the polycrystalline mass was used as the charge for vertical Bridgman growth (Figure 2\emph{a}). For the vertical growth, a quartz crucible (1.3 cm in diameter) was placed in the $\approx$$820\,^{\circ}\mathrm{C}$ region of the furnace, where it was then translated at a rate of 0.2 mm/h through a temperature gradient of $10\,^{\circ}\mathrm{C}$/cm.  Figure 2\emph{b} shows single crystal grains for a Ga$_2$(Se$_{0.2}$Te$_{0.8}$)$_3$ specimen obtained from the growth. It is important to note that an immiscibility gap in the range of selenium fractions 0.5$\leq$\emph{x}$\leq$0.9 has been observed, thus limiting Ga$_2$(Se$_x$Te$_{1-x}$)$_3$ to the tellurium rich condition. Such a phenomenon has been previously reported by Warren et. al. \cite{Warren_1974}    

\begin{figure}[h]
\includegraphics[width=8cm]{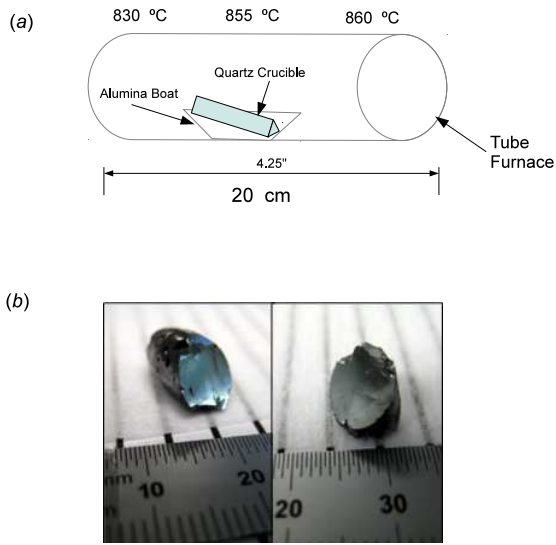}
\caption{(\emph{a}) Schematic showing the temperature profile and crystal growth setup using a horizontal tube furnace. (\emph{b}) Single crystal grains of Ga$_2$(Se$_{0.1}$Te$_{0.9}$)$_3$ with size on the order of 3 to 5 mm.}
\label{fig1}
\end{figure}

\begin{figure}[h]
\includegraphics[width=8cm]{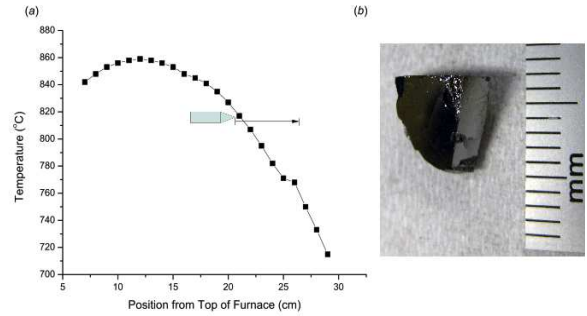}
\caption{(\emph{a}) Temperature profile of the vertical Bridgman furnace used for growing Ga$_2$(Se$_x$Te$_{1-x}$)$_3$ specimens. The crucible was translated vertically out of the $\approx$$820\,^{\circ}\mathrm{C}$ region at a rate of 0.2 mm/h. (\emph{b}) Ga$_2$(Se$_{0.2}$Te$_{0.8}$)$_3$ single crystal grains on the order of 1 cm obtained from the growth.}  
\label{fig2a}
\end{figure} 

\section{Structure Characterization}

Structure and phase verification was carried via high-resolution powder x-ray diffractometry. The measurements were carried out at the Advanced Photon Source (APS) at Argonne National Laboratory (ANL). Figure 3 shows the diffraction pattern for a Ga$_2$(Se$_{0.3}$Te$_{0.7}$)$_3$ specimen with a Rietveld refinement implemented. A cubic zincblende structure was confirmed with a lattice constant of $\approx$5.8 $\AA$ and a computed density of $\approx$5.4 g/cm$^3$. The inset of Figure 3 highlights the presence of satellite peaks around the (111) Bragg line not associated with crystal lattice reflections. Such an occurrence has been previously observed for annealed specimens of the binary compound Ga$_2$Te$_3$ \cite{Kurosaki_2008, Kim_2011} . High resolution electron microscopy images of Ga$_2$Te$_3$ have shown the presence of vacancy planes along the [111] direction occurring periodically at 3.5 nm intervals\cite{Kurosaki_2008, Kim_2011}. Examining the two pairs of satellite peaks in the Ga$_2$(Se$_{0.3}$Te$_{0.7}$)$_3$ diffraction pattern, the intensities of the peaks on opposites sides of the (111) Bragg line are not commensurate with each other. This seems indicative of only short range ordering due to presence of a distortion possibly arising from the shared nature of the anion (Te/Se) site in the ternary Ga$_2$(Se$_x$Te$_{1-x}$)$_3$ system.            

\begin{figure}[htb]
\includegraphics[width=8cm]{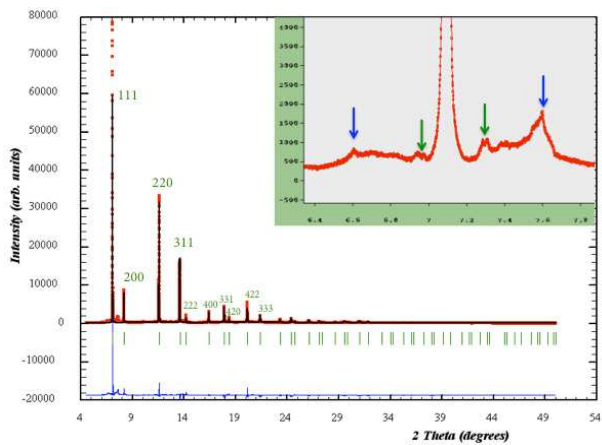}
\caption{High resolution diffraction pattern of Ga$_2$(Se$_{0.3}$Te$_{0.7}$)$_3$ obtained using synchrotron x-rays at the APS 11-BM beam line at ANL. The red points represent the experimental data, the black line represents the simulated pattern, the blue line represents the error between experiment and calculation, and the green represents all possible Bragg reflections for a zincblende lattice. The inset shows two pairs of satellite peaks, depicted by the green and blue arrows, around the (111) Bragg line.}
\label{fig3}
\end{figure}

\section{Defect Studies}

Positron lifetime and coincidence Doppler broadening (\emph{i.e.} electron momentum) measurements were carried out on Ga$_2$(Se$_x$Te$_{1-x}$)$_3$ samples of selenium fraction, \emph{x}, ranging from 0.0 to 0.3. Measuring positron lifetime and sample electron momentum allows one to identify defects and probe their sizes/concentrations. The specific details of carrying out such measurements have been previously reported \cite{Krause_1999, Lynn_1977, Saito_2002}. A $^{22}$Na source is used to inject positrons into the sample. Lifetime measurements involve a coincidence setup of two BaF$_2$ detectors to measure the time difference between the 1.27 MeV de-excitation gamma ray coming from the $^{22}$Na source, and the 511 keV gamma ray originating from positron annihilation in the sample. Coincidence Doppler broadening measurements observe the red/blue shifts of the annihilation photons using Ge detectors; with this information the electron momentum can then be probed via energy and momentum conservation.

Figure 4 shows the average positron lifetime (computed by fitting a two component lifetime model to the experimental positron decay spectrum) as a function of selenium atomic fraction for four Ga$_2$(Se$_x$Te$_{1-x}$)$_3$ samples. Positrons in all of the samples had an average lifetime of $\approx$350 to 400 ps, indicative of a strong presence of defects (bulk annihilation lifetime for common semiconductors is on the order of $\approx$200 ps \cite{Krause_1999}). This is also evinced by the observation that all of the samples had a dominant second lifetime component (nominally associated with defects), with the exception of the sample with \emph{x} = 0.3 which had a fairly large 347 ps first lifetime component (possibly related to a highly prominent defect).

\begin{figure}[htb]
\includegraphics[width=8cm]{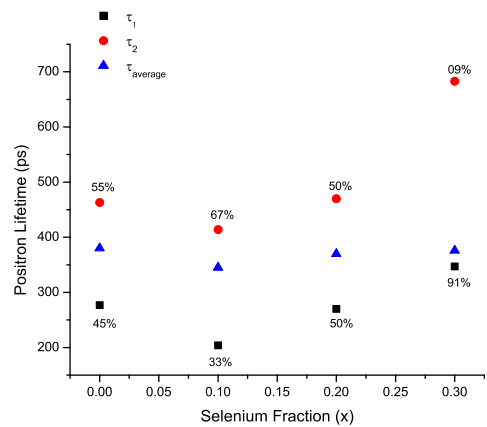}
\caption{Positron lifetimes (computed via a two-lifetime component model) as a function selenium atom fraction with the intensities of each lifetime component. The first lifetime component is denoted by $\tau$$_1$, the second lifetime component by $\tau$$_2$, and the average lifetime component by $\tau$$_{average}$.}  
\label{fig4}
\end{figure}

Analyzing the electron momentum relative to the electron momentum distribution measured for elemental tellurium (Figure 5) can give insight on the type of defect that is dominant in the Ga$_2$(Se$_x$Te$_{1-x}$)$_3$ samples. Large intensities of low electron momenta (0 to $\approx$0.03\emph{m}$_e$\emph{c}) are observed (\emph{m}$_e$ is the electron rest mass and \emph{c} is the speed of light); after $\approx$0.03\emph{m}$_e$\emph{c} electron momentum intensities drop. A similar trend is observed when the data are normalized relative to elemental gallium and selenium. Such behavior suggests that positrons are not annihilating at single gallium, tellurium, or selenium vacancies, but at large open-volume defects such as vacancy clusters or voids where the overall electron density is low and low-momentum valence electrons dominate.

\begin{figure}[htb]
\includegraphics[width=8cm]{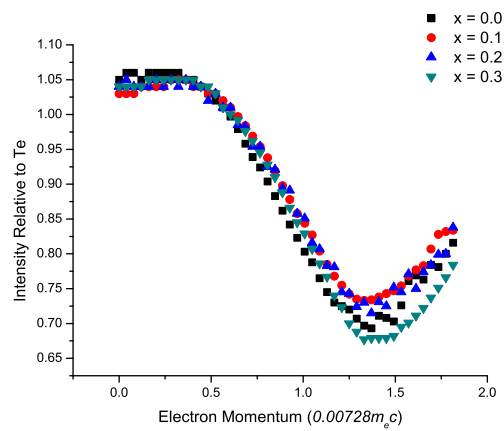}
\caption{Electron momentum distributions for Ga$_2$(Se$_x$Te$_{1-x}$)$_3$ samples with selenium atomic fraction ranging from 0.0 to 0.3.}  
\label{fig5}
\end{figure}

\section{Conclusions}

Single crystals of Ga$_2$(Se$_x$Te$_{1-x}$)$_3$, for \emph{x} = 0.1, 0.2, 0.3, semiconductors have been obtained via modified Bridgman growth methods. High-resolution powder x-ray diffractometry confirmed a cubic zincblende structure. The presence of satellite peaks around the (111) Bragg line suggests an additional presence of short-range structural vacancy ordering. Preliminary positron annihilation spectroscopy measurements on resulting as-grown specimens indicate a proclivity for vacancy clustering and/or void formation.
\section{Acknowledgements}        

The authors would like to acknowledge Christopher Ramsey, Zewu Yan, Gautam Gundiah, Donghua Xu, and the 11-BM beamline staff at ANL for their assistance in conducting this work.. This research was performed under appointment to the U.S. Department of Energy Nuclear Nonproliferation International Safeguards Graduate Fellowship Program sponsored by the National Nuclear Security Administration Office of Nonproliferation and International Security. This work was supported by the U.S. Department of Energy/NNSA/NA22 and carried out at Lawrence Berkeley National Laboratory under Contract NO. AC02-05CH11231. Work at the University of California, Berkeley is supported by the DHS-NSF Academic Research Initiative Grant.

\end{document}